\documentclass[aps,prl,twocolumn,superscriptaddress]{revtex4}
\usepackage{mathrsfs}
\usepackage{epsfig}
\usepackage{graphicx}
\usepackage{amsfonts}
\usepackage[figuresright]{rotating}
\usepackage{amssymb}
\usepackage{amsmath}
\usepackage{dcolumn}
\usepackage{bm}
\usepackage{color}

\newcommand{\lmu} {Department of Physics and Arnold Sommerfeld Center for Theoretical Physics,
Ludwig-Maximilians-Universit{\"a}t M{\"u}nchen, Theresienstr.\ 37,
80333 Munich, Germany}

\begin{document}
\title{Accessing many-body localized states through the Generalized Gibbs Ensemble }

\author{Stephen Inglis}
\affiliation{\lmu}

\author{Lode Pollet}
\affiliation{\lmu}

\begin{abstract}
We show how the thermodynamic properties of large many-body localized systems can be studied using quantum Monte Carlo simulations. To this end we devise a heuristic way of constructing local integrals of motion of very high quality, which are added to the Hamiltonian in conjunction with Lagrange multipliers. The ground state simulation of the shifted Hamiltonian corresponds to a high-energy state of the original Hamiltonian in case of exactly known local integrals of motion. We can show that the inevitable mixing between eigenstates as a consequence of non-perfect integrals of motion is weak enough such that the characteristics of many-body localized systems are not averaged out in our approach, unlike the standard ensembles of statistical mechanics. Our method paves the way to study higher dimensions  and indicates that a full many-body localized phase in 2d, where (nearly) all eigenstates are localized, is likely to exist.
\end{abstract}


\maketitle

{\it Introduction --}
Many-body localization (MBL) addresses the fundamental question under which conditions quantum systems can avoid ergodicity and thermalization, thereby generalizing Anderson localization to interacting systems~\cite{Anderson1958, Mikheev1983, KaganMaksimov, Basko2006, MirlinPolyakov2005, Nandkishore_review2015, Altman_review2015}. The widely accepted mechanism for thermalization in quantum systems is the eigenstate thermalization hypothesis~\cite{Deutsch91, Srednicki94, Rigol08}: Under very mild assumptions, (almost) every eigenstate of the system is thermal. This implies that the reduced density matrix of a small subsystem, obtained by tracing out the degrees of freedom of the considered (eigen)state outside the subsystem, is indistinguishable from the thermal density matrix with an effective temperature that depends on the energy density of the chosen eigenstate.
MBL states, on the contrary, retain knowledge of their initial local conditions in local operators for asymptotically large times. The picture of local integrals of motion (LIOM)~\cite{Serbyn2013, Vosk2013, Huse2014, Ros2015, Chandran2015, Imbrie2014} can explain most of the unusual phenomenology of MBL states: an area low holds in space leading to a logarithmic growth of the entanglement entropy in time and space, and the dc conductivity is identically zero. The area law was demonstrated explicitly in Ref.~\cite{BauerNayak2013}, while it was also found that rare regions can lead to deviations. Dynamics is a decisive characteristic to distinguish between thermal and MBL states: experiments on trapped ions~\cite{Smith2015} and 1d cold atoms~\cite{Schreiber2015} demonstrated memory of the initial conditions over long periods of time for sufficiently strong disorder.  \\

The thermodynamic predictions of a single-eigenstate ensemble are fundamentally different from the ones of the standard (micro-/grand-)canonical ensembles of statistical mechanics in the MBL phase.  Obtaining a single, or even a certain class of eigenstates in the middle of the spectrum in order to prevent the averaging over all states with the same energy, which inevitably occurs in the standard ensembles,  is a daunting task from a computational point of view because the mean level spacing decreases exponentially with system size. If one wants to study sufficiently large MBL systems, say of size at least $100$ in 1d (which is just $10 \times 10$ in 2d), the mean level spacing is far below the machine precision. So far exact diagonalization (ED) for disordered spin-1/2 systems could be performed for 22 sites using a shift-invert method~\cite{Luitz2015}. 
With Density Matrix Renormalization group (DMRG) methods  one could go to larger system sizes by using energy projection~\cite{PekkerClark2015, Karrasch2015, Sheng2015} according to $(\hat{H} - \sigma)^2$, with $\sigma$ the target energy, or by using the successful shift invert  $(\hat{H} - \sigma)^{-1}$ method~\cite{PekkerClark2015}, or performing the matrix product selection on the basis of explicit LIOMs~\cite{PollmannSondhi2015}.


In this Letter we show how quantum Monte Carlo methods can be used to study the equilibrium properties of MBL states provided the picture of LIOMs hold, which seems to be acceptable (deep) in the fMBL (full-MBL) phase, where (nearly) all many-body eigenstates are localized~\cite{Pekker2014a, Pekker2014b, Chandran2014, PollmanCirac2015}. The key idea is that, if by some technique a finite density of LIOMs can be constructed, they can be added to the action in conjunction with Lagrange multipliers. This creates a generalized Gibbs ensemble (GGE), which allows to map MBL states to ground states, amenable to quantum Monte Carlo simulations for sign-free models. We provide a heuristic way of constructing LIOMs, whose quality can be tested a posteriori. Our method allows to distinguish between ergodic and non-ergodic phases, at least sufficiently far away from the transition. Only in the MBL phase can the properties of high-energy states be studied; we focus in particular on the eigenvalues and eigenvectors of the reduced one-body density matrix~\cite{Bera2015}, which we apply here to hard-core bosonic  instead of fermionic systems.

{\it Method --}
Our first objective is to generate a set of LIOMs (which are by no means unique) of sufficiently good quality such that the spectrum is (nearly) unaffected when adding them  to the action in conjunction with Lagrange multipliers. A number of procedures have been suggested in the literature, including infinite time evolution~\cite{Chandran2015}, self-similar transformations~\cite{Rademaker2016}, renormalization methods~\cite{Monthus2016} and perturbative approaches~\cite{Ros2015}. Although these approaches could also be applied here, we suggest another approach with the purpose of obtaining the LIOMs in a simple enough operator formulation, namely one that is compatible with existing quantum Monte-Carlo worm-type algorithms~\cite{Prokofev1998} in the implementation of Ref.~\cite{Pollet2007}, {\it i.e.}, we look for LIOMs in the form of L-bit operators $\mathbf{L}$, defined over a finite support $S$, which contain operators  that are only of nearest-neighbor hopping, local density or nearest-neighbor density-density type. Specifically, in 1d they take the form
\begin{align}
\mathbf{L}^{(p)} = \sum_{i \in S} \mu_i^{(p)} n_i + \sum_{i, i+1 \in S} t_i^{(p)} (b_{i+1}^\dagger b_i + h.c.) + V_i^{(p)} n_i n_{i+1},
\label{eq:Lbit}
\end{align}
The coefficients $\mu_i^{(p)}, t_i^{(p)}$, and $V_i^{(p)}$ are optimization parameters for the L-bit defined on patch $p$. The support $S$ is a strict subset of the sites $p$, $S \subset p$. We take the patches to be non-overlapping spatial sections of the full Hamiltonian. We seek to find  the single best-possible L-bit of the above form over the patch $p$ in the sense that the norm of the commutator with the Hamiltonian over this patch is as small as possible. In a fMBL system the support of the L-bits is exponential in the localization length and the proximity of the atomic limit suggests that Eq.~\eqref{eq:Lbit} is a good parametrization. L-bits defined over different patches have an exponentially small overlap. Note that our design consists of finding some L-bits (not all possible ones) which in general have a spectrum of more than 2 states. Other design criteria are certainly possible but left for future work.

The heuristic optimization proceeds as follows using full diagonalization over the full patch. We enforce that the width of the L-bit spectrum be equal to 1 in order to meaningfully compare the quality of the proposed L-bits. We rotate the L-bit to the eigenbasis of the patch Hamiltonian, $\tilde{\mathbf{L}} = \mathbf{U}^\dagger \mathbf{L} \mathbf{U}$ with $U$ the matrix of eigenvectors of the patch Hamiltonian. In the case that $ [ \mathbf{L}^{(p)}, H^{(p)}] =0$, $\tilde{\mathbf{L}}^{(p)}$ is a diagonal matrix. Keeping this in mind, we define the cost function as the Frobenius norm of the off-diagonal elements of matrix $\tilde{\mathbf{L}}^{(p)}$. A non-linear minimization solver is used to find the optimal parameters $\mu_i^{(p)}, t_i^{(p)}$, and $V_i^{(p)}$. Other ways to construct the L-bits can certainly be thought of (and avoid the full diagonalization step over the patch), and may perhaps be better suited to combine with tensor network state methods or yield a smaller value for the commutator; our goal is merely to show that  using the GGE is realistic and robust.

To check whether the operators thus found are L-bits of good enough quality, we shift the original Hamiltonian with the optimal L-bits found over the $N_p$ patches and examine the ground state of the GGE Hamiltonian
\begin{align}
\mathbf{H}_G(\lambda_1, \ldots \lambda_{N_p}) = \mathbf{H} + \sum_{p=1}^{N_p} \lambda_p \mathbf{L}^{(p)}. \label{eq:gge}
\end{align}
We investigate each of the $\left< \mathbf{L}^{(p)} \right>$ as a function of $\lambda_p$ in a quantum Monte Carlo simulation. Our tests (also performed for the Aubry-Andre model) show that the value of the  $\left< \mathbf{L}^{(p)} \right>$'s can, to a very good accuracy, be set by the respective $\lambda_p$ independenly of the L-bits on the other patches. If successful, $\left< \mathbf{L}^{(p)} \right>$ shows large plateaus separated by sharp jumps, {\it i.e.}, the L-bit selects the ground state based on the expectation of the L-bit for that state without mixing the states.
When no plateaus can be found, the L-bit must be discarded. If no L-bits at all can be found, then the procedure failed and the system is certainly ergodic. Deviations from a flat pleateau are an indication of how much hybridization took place as a consequence of non-ideal L-bits and the truncation of their exponential tails. This averaging is however tolerable because it averages low energy states within the GGE (which are all within a given L-bit sector), in contrast to the (mirco)-canonical averaging which averages over all orientations of the L-operators within a narrow energy window. In the MBL phase one can move states around in the spectrum by choosing the $\lambda_p$ according to different plateau values. The difference in energy between the plateaus is proportional to the energy gain in the many-body spectrum. Note that, in contrast to textbook Legendre transformations, our L-bits are intensive operators and we need a finite density of them in order to reach high-energy states.

\begin{figure}
\includegraphics[width=0.80\columnwidth]{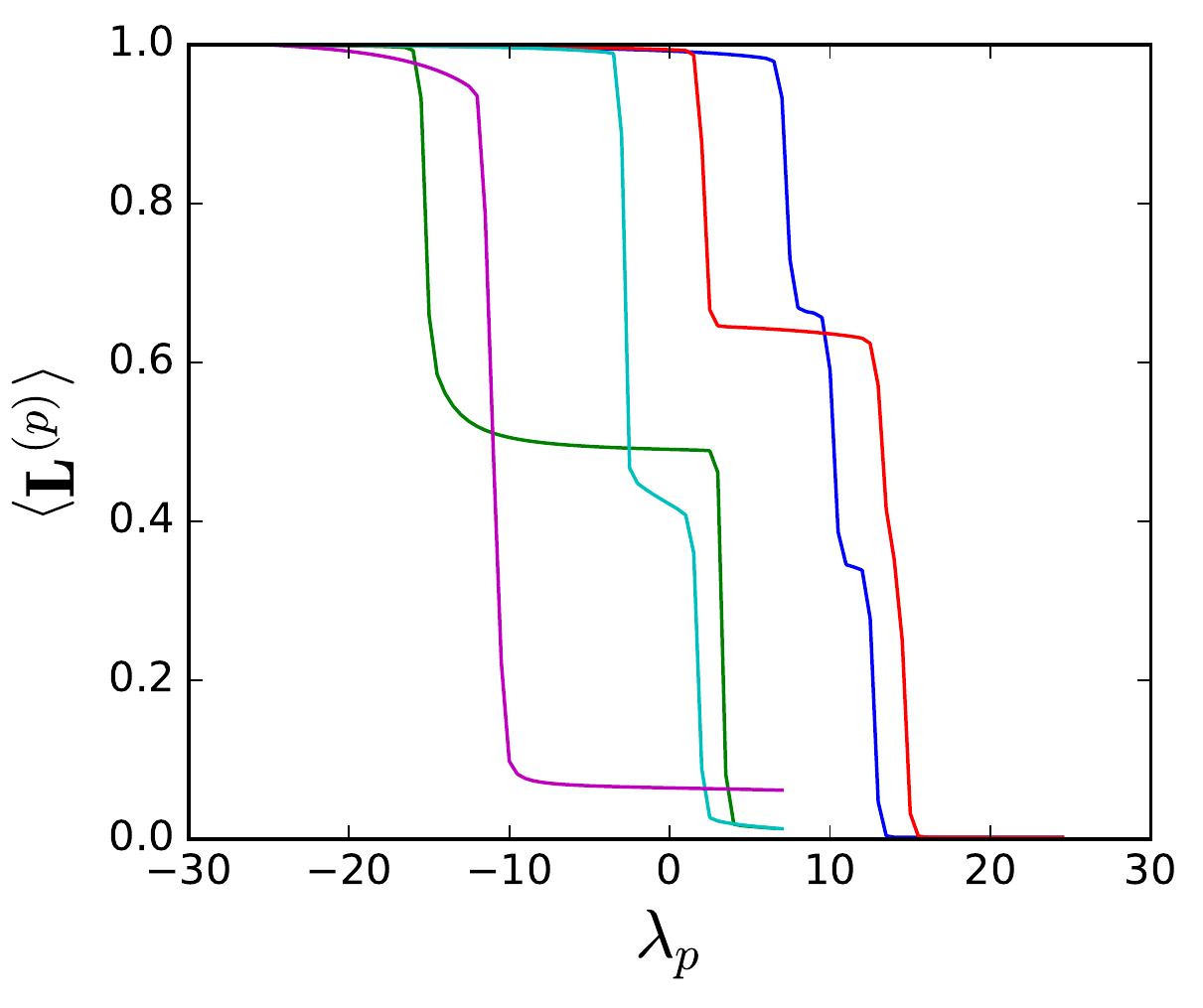}
\caption{(Color online). Expectation value of 5 different L-bits with 5 independent $\lambda$ for $\Delta=6$ on a support of 3 sites in a patch of 11 sites for the 1d Heisenberg model in a random magnetic field. Distinct plateaus are seen for each $\left< \mathbf{L}^{(p)} \right>$ demonstrating that the L-bits represent well-defined quantities.
\label{fig:plateau_MBL}}
\end{figure}

\begin{figure}
\includegraphics[width=0.80\columnwidth]{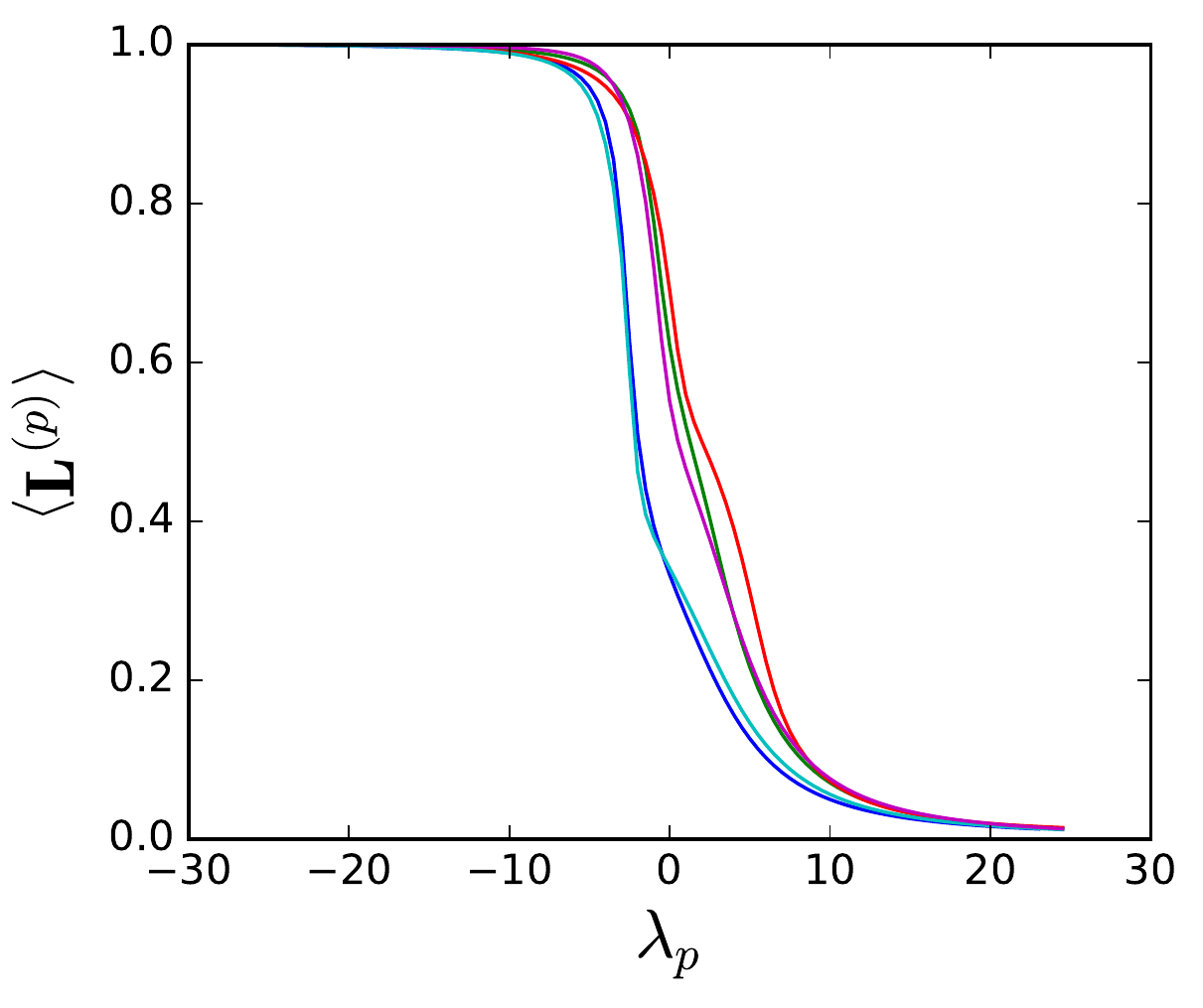}
\caption{(Color online). Expectation value of 5 distinct L-bits for $\Delta=1$ on a support of 3 sites in a patch of 11 sites for the 1d Heisenberg model in a random magnetic field. No plateaus are seen in the range $\lambda \in ]-10,10[$; all attempted L-bits are hence discarded.
\label{fig:plateau_ergodic}}
\end{figure}

{\it Results in 1d --}
We now apply this construction to the 1d spin-1/2 Heisenberg chain with disorder in the magnetic field. For convenience, we write the Hamiltonian in the hard-core boson language as  $H = -t\sum_i b_i^{\dagger}b_{i+1} + 2t \sum_i n_{i+1} n_i + \sum_i \mu_in_i$, where $\mu_i$ is drawn uniformly from $[-\Delta, \Delta]$ and the hopping amplitude is set to one, $t=1$. 
The system is fMBL for $\Delta > 3.5$ according to the full diagonalization results of Ref.~\cite{Luitz2015} (However, Ref.~\cite{Prelovsek2016} shows that truly zero dc conductivity may require much larger disorder strength).
First, we compare the quality of the plateaus as a function of the value of the corresponding Lagrange multiplier in the fMBL phase (Fig.~\ref{fig:plateau_MBL}) and in the ergodic phase  (Fig.~\ref{fig:plateau_ergodic}). In the fMBL phase, sharp plateaus are observed indicating that the respective L-bits (nearly) commute with the Hamiltonian. For $\lambda_p = 0$ we always see a well-defined plateau, corresponding to the localized ground state of the original Hamiltonian. By contrast, in the ergodic phase no clear plateaus can be discerned, which is also obvious from the strong hybridization at $\lambda_p = 0$.

To judge the quality of the L-bits, we define a measure defined as
\begin{align}
Q = \left( \sum_k^n \left< \mathbf{L}(\lambda^k) \right> - \left< \mathbf{L}(\lambda^{k+1}) \right> \right)^2
\label{eq:quality}
\end{align}
where $\lambda^k$ are a mesh of points amenable to calculating using quantum Monte Carlo.
The number of $\lambda^k$ points chosen is fixed for every L-bit, and the spacing is determined by the energy density of the ground state of the unmodified Hamiltonian.
This is because, typically, the magnitude of $\lambda$ needed to change $\left< \mathbf{L} \right>$ scales with this quantity.
This measure is inspired by the inverse participation ratio, and has a maximal value of 1 in the case of a single large jump of size 1 (the width of the L-bit spectrum) separating two perfectly flat plateaus.

\begin{figure}
\includegraphics[width=0.80\columnwidth]{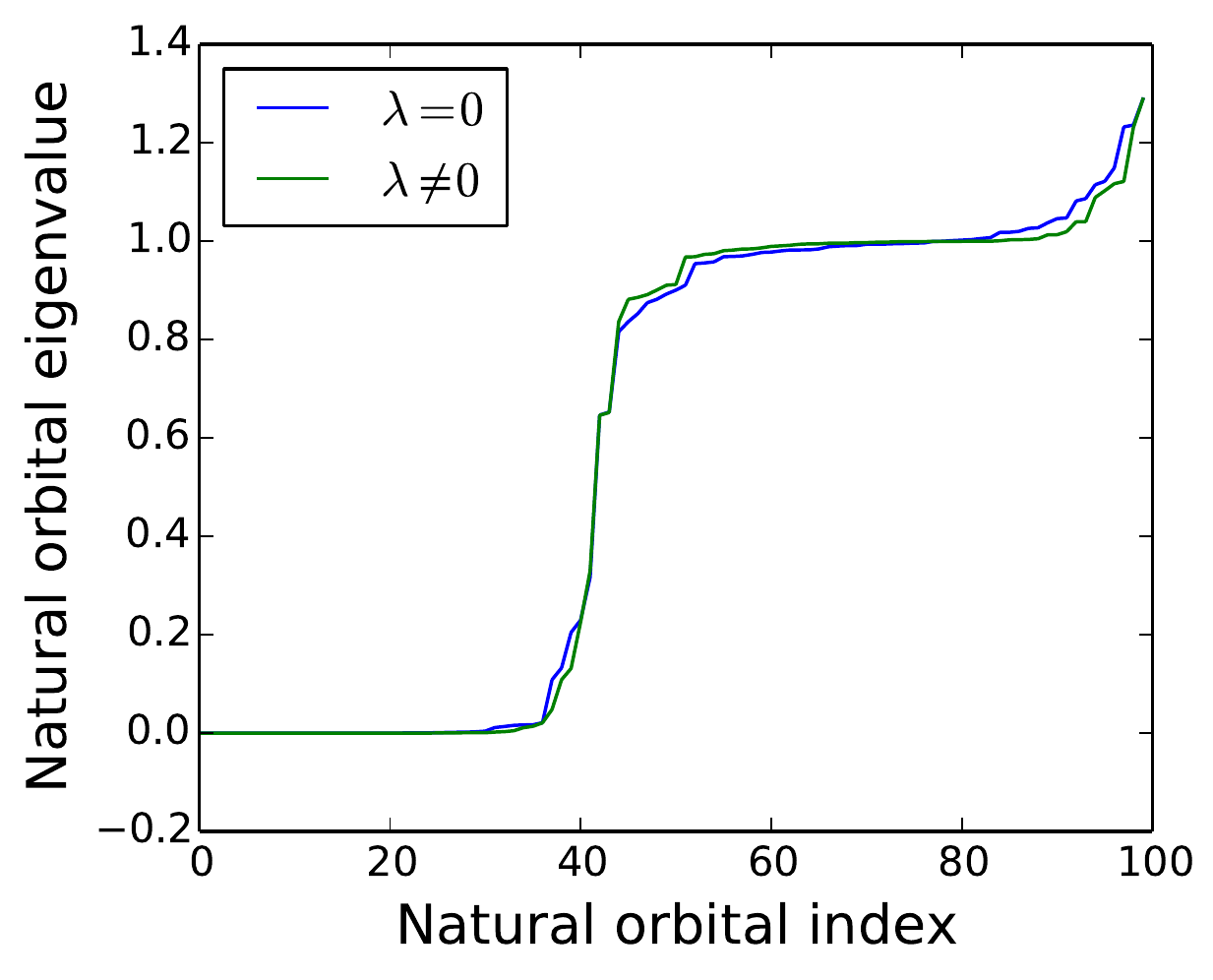}
\caption{(Color online). Comparison between the sorted eigenvalues corresponding to the natural orbitals in the ground state and a MBL state for the 1d Heisenberg model for $\Delta = 6$. We constructed 12 L-bits on a system of size 100. The ground state energy is $E/J = -180.67(1)$ and the energy of the excited state is $E/J = -135.98(1)$.
\label{fig:1d_d6_no}}
\end{figure}

Second, when good plateaus are found, we invoke transitions by selecting values of the Lagrange multipliers corresponding to plateaus found for $\lambda_p \neq 0$. The quantum Monte Carlo simulation projects on the ground state of the shifted Hamiltonian and gives us access to its thermodynamic properties, including the single particle density matrix, from which we can extract the localization length and which we diagonalize in order to obtain the natural orbitals and their corresponding eigenvalues. According to Ref.~\cite{Bera2015}, the sorted eigenvalues should show a sharp jump in the MBL phase for a fermionic system. This is also what we see in Fig.~\ref{fig:1d_d6_no}: the jump seen for the ground state of the GGE Hamiltonian is very similar way to the jump seen for the ground state of the original Hamiltonian, which we certainly expect to hold for small localization lengths.
The energy increase we could reach in this example is about $25\%$ of the original spectrum (we did not attempt to go higher). Furthermore, we note that the winding number remains zero and that our results for local quantities remain invariant when increasing the system size. Given the current state-of-the art of quantum Monte Carlo simulations with worm-type updates, fMBL properties of systems of several thousands of sites are straightforward with this approach. \\

\begin{figure}
\includegraphics[width=0.80\columnwidth]{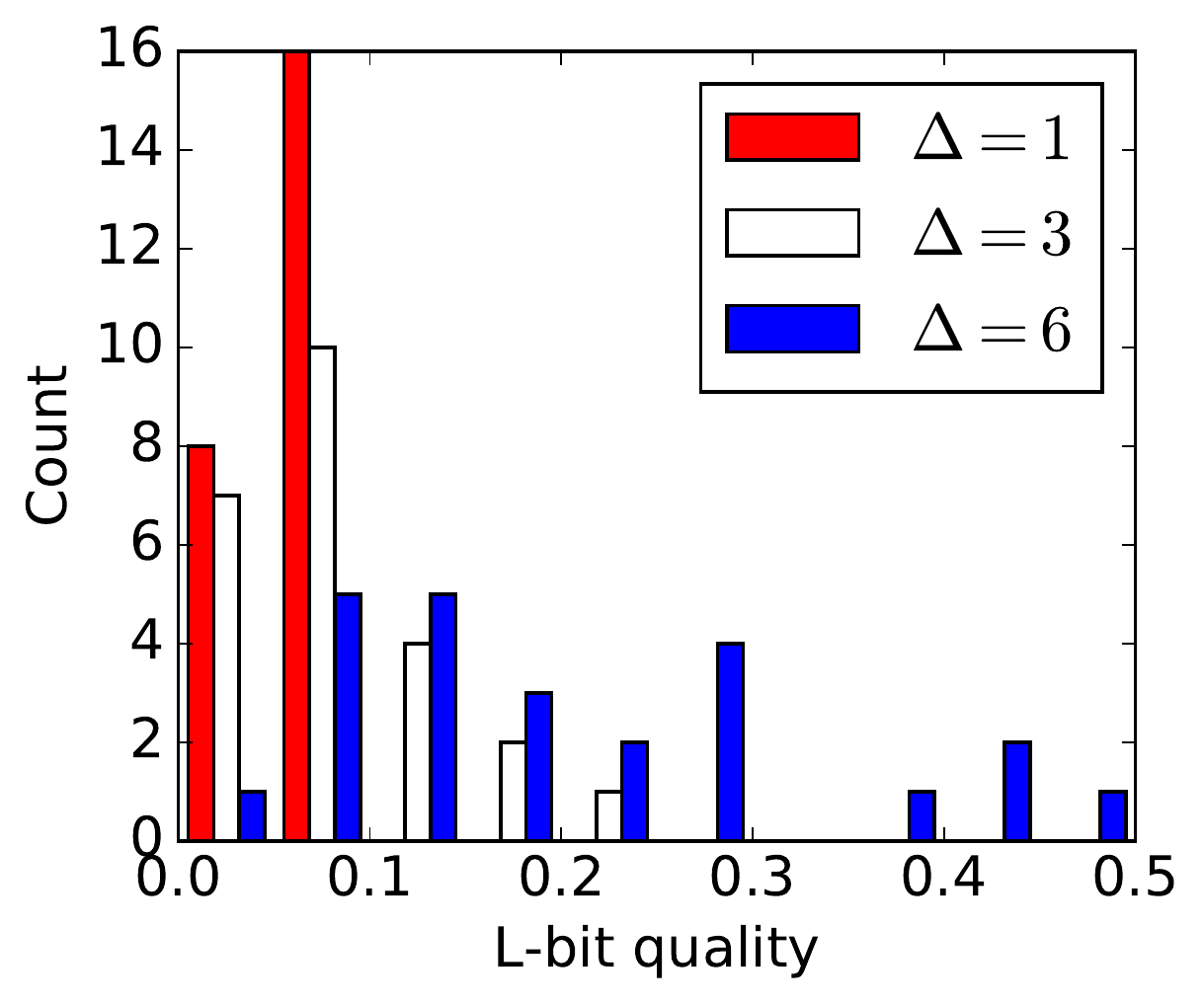}
\caption{(Color online). Histogram of the quality of the L-bit for different disorder strength using Eq.~\eqref{eq:quality}.
\label{fig:lambda_cost}}
\end{figure}

When we apply the same procedure to intermediate disorder strengths $\Delta \sim 3$, where the exact diagonalization results of Ref.~\cite{Luitz2015} found a mobility edge (which is however contentious~\cite{DeRoeck2016}), we observe that we can still find good L-bits but considerably fewer than in the fMBL phase. This is shown in Fig.~\ref{fig:lambda_cost} where we see that the L-bits are always of poor quality in the ergodic phase (and must hence be discarded), almost always of good quality in the fMBL phase, whereas in the intermediate regime we get a broad distribution. Discarding the poor L-bits but making use of the high-quality ones, we could go up about $15\%$ in energy for 9 L-bits over a system of size 100 for $\Delta = 3$. The winding number and the properties of the natural orbitals indicate insulating behavior. These results suggest that certain states are localized while others are delocalized in the spectrum.

\begin{figure}
\includegraphics[width=0.80\columnwidth]{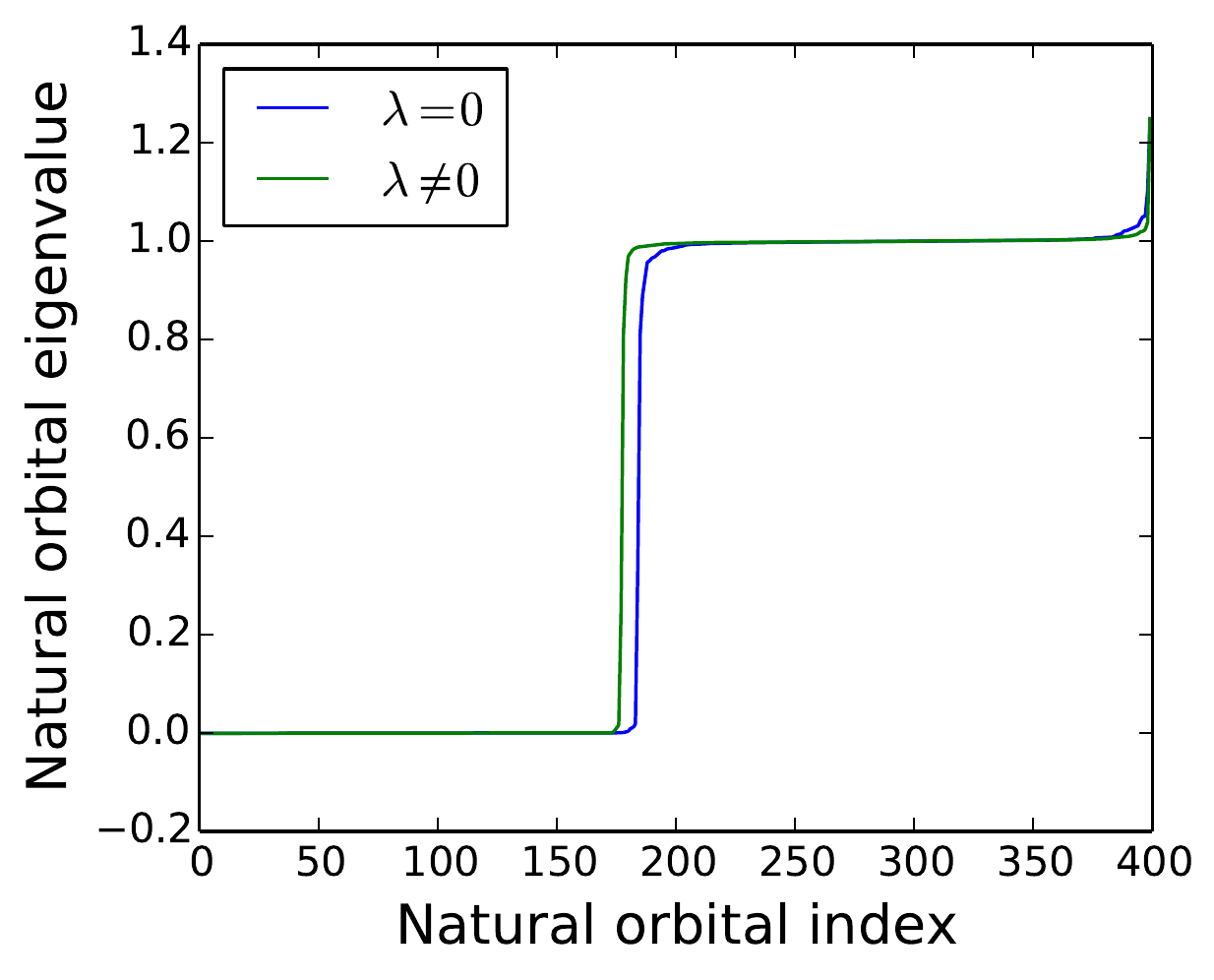}
\caption{(Color online). Sorted eigenvalues of the natural orbitals for the 2d Heisenberg model in the ground state and for a system with 40 L-bits obtained by quantum Monte Carlo simulation of size $20 \times 20$. The L-bits may change the particle number compared to the ground state, causing the shift in the location of the jump.
\label{fig:2d_NO}}
\end{figure}

\begin{figure}
\includegraphics[width=0.80\columnwidth]{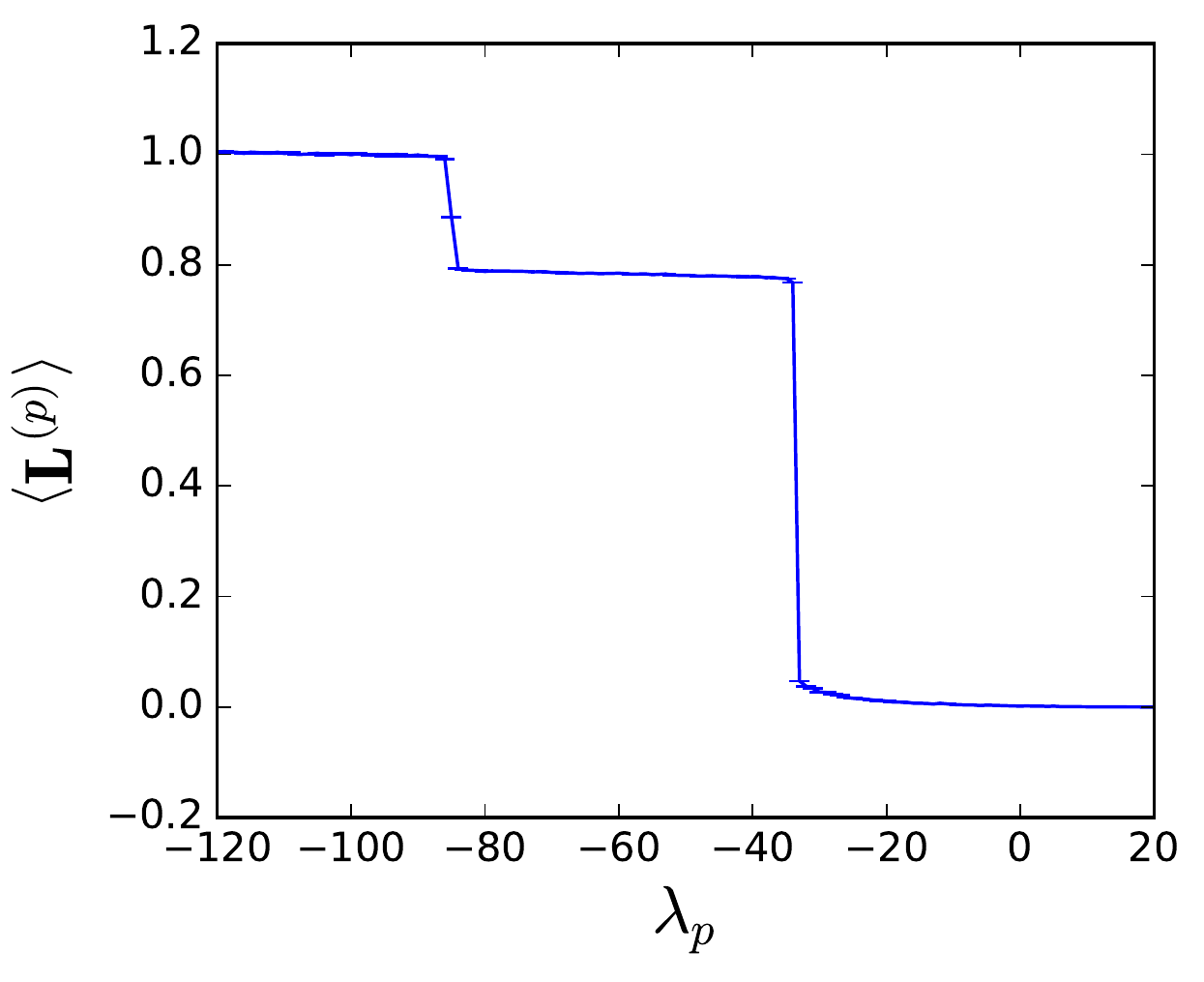}
\caption{(Color online). Illustration of the plateaus found for a single $\mathbf{L}$ as a function of $\lambda$ for a 4-site L-bit embedded in a two dimensional $20\times20$ lattice with $\Delta=40$.
With large $\Delta$ effective L-bits can be optimized even when the embedding Hamiltonian is small.
\label{fig:2d_plateau}}
\end{figure}

{\it Results in 2d --} The most appealing feature of a quantum Monte Carlo approach is the possibility to study the static properties of MBL states in higher dimensions. Here we demonstrate that the L-bit construction can be done in two (and generally higher) dimensions as well. As a model we take the 2d generalization of the disordered 1d Heisenberg model introduced before.
Deep in the fMBL phase (assuming it exists) we take for the support a $2 \times 2$ cluster which is diagonalized over a 12 site patch (a $4 \times 4$ square excluding the 4 corners), and allow the same site- and bond-operators as in the 1d case. To test the quality of the L-bits, we embed them in a $20\times20$ lattice and measure the plateaus using quantum Monte Carlo simulations.
For $\Delta/J=40$ we are able to find L-bits with good plateaus, as shown in Fig.~\ref{fig:2d_plateau}. Also the sorted eigenvalues of the natural orbitals of the system with an energy about $25\%$ (40 L-bits were constructed) above the ground state show a pronounced jump similar as in the ground state. Fig.~\ref{fig:2d_NO} demonstrates how the addition of the L-bits in 2d does not affect the natural orbital structure suggesting that fMBL can also be realized in 2d~\cite{BarLev}.

{\it Conclusions --}
We have demonstrated the ability to find and implement L-bits, effectively allowing us to access excited states of a model as the ground states of a modified Hamiltonian which is related by a Legendre transform to the original model. When the L-bits do not exactly commute with the Hamiltonian, mixing occurs in the original Hamiltonian, but this does not destroy the MBL properties thanks to the GGE ensemble and the fact that local observables can distinguish between nearby eigenstates in the MBL phase. The system sizes that could be reached are substantially larger than with other methods, and the approach opens the way to study fMBL in dimensions higher than one.  The GGE approach can also be combined with DMRG where more complicated operators in the L-bit construction can be taken into account, as long as the localization length remains sufficiently small. 
With a basic framework in place, there are a few natural extensions that can be explored in future work. It would be interesting to use a cost function which does not involve exact diagonalization over the patch, such as for instance DMRG or a stochastic optimization based on the quality of the plateaus as a function of the Lagrange multiplier, which could all be highly parallelized. Our approach relies crucially on the existence of the LIOMs and as such it is not clear whether it can be used to study the transition (or perhaps crossover~\cite{Prelovsek2016}). Properties such as the entanglement entropy (in higher dimensions) and observing resonances, or studying the particle-hole symmetric case, would be interesting.

{\it Note -- }
During the final stages of this work, a cold-atom experiment showed the existence of a MBL phase in two dimensions~\cite{Bloch_MBL_2d}.

{\it Acknowledgments --}
The authors would like to acknowledge inspiring discussions with E. Altman, B. Bauer, E. Demler, F. Heidrich-Meisner, D. Huse, D. Luitz, D. Pekker, N. Prokof'ev, F. Pollmann, and U. Schneider. This work was supported by FP7/ERC starting grant No. 306897 and FP7/Marie-Curie CIG Grant No. 321918. Part of the work of LP was performed at the Aspen Center for Physics, which is supported by National Science Foundation grant PHY-1066293

\bibliography{refs}{}

\end{document}